\documentclass[
10pt,a4paper,twocolumn,aps,prx, superscriptaddress,longbibliography, nofootinbib, amsmath,amssymb,nobalancelastpage
]{revtex4-2}

\usepackage{etoolbox}
\usepackage{mathtools}
\usepackage{graphicx}
\usepackage{dcolumn}
\usepackage{bm}
\usepackage[svgnames]{xcolor}
\usepackage{siunitx}
\usepackage{booktabs} 
\usepackage{array}
\usepackage{braket}
\usepackage{microtype}
\usepackage{verbatim}
\usepackage{lmodern}

\usepackage[colorlinks=true, linkcolor=blue, citecolor=DarkGreen, urlcolor=orange, pdfauthor={Author}, pdftitle={Title}]{hyperref}

\newcommand{\gtilde}{\tilde{g}}
\newcommand{\gtildeA}{\tilde{g}_{\rm A}}

\newcommand{\gtildeAB}{\tilde{g}_{\rm A/B}}
\newcommand{\kappaA}{\kappa_{\rm A}}
\newcommand{\kappaB}{\kappa_{\rm B}}
\newcommand{\kappaAB}{\kappa_{\rm A/B}}
\newcommand{\sech}{{\rm sech}}
\newcommand{\DeltaAB}{\Delta_{\rm AB}}
\newcommand{\deltaA}{\delta_{\rm A}}
\newcommand{\deltaB}{\delta_{\rm B}}
\newcommand{\deltaAB}{\delta_{\rm A/B}}
\newcommand{\nA}{{n_{\rm A}}}
\newcommand{\nB}{{n_{\rm B}}}

\usepackage{xparse}
\usepackage{hyperref}
\ExplSyntaxOn
\NewDocumentCommand{\figref}{m o}
{
  \IfNoValueTF{#2}
    {
      Fig.~\hyperref[#1]{\ref*{#1}}
    }
    {
      \tl_if_single:nTF {#2}
        {Fig.~\hyperref[#1]{\ref*{#1}(#2)}}
        {Figs.~\hyperref[#1]{\ref*{#1}(#2)}}
    }
}
\ExplSyntaxOff

\makeatletter
\renewcommand{\bibsection}{
  \par
  \baselineskip26\p@
  \bib@device{\linewidth}{82\p@}
  \nobreak\@nobreaktrue
  \addvspace{19\p@}
  \par
}
\makeatother

\begin{document}

\title{Emission and Absorption of Microwave Photons in Orthogonal Temporal Modes\\
across a 30-Meter Two-Node Network}

\author{Alonso Hernández-Antón}\email{Contact author: ahernandez@phys.ethz.ch}
\affiliation{Department of Physics, ETH Zurich, CH-8093 Zurich, Switzerland}
\affiliation{Quantum Center, ETH Zurich, CH-8093 Zurich, Switzerland}

\author{Josua D. Schär}
\affiliation{Department of Physics, ETH Zurich, CH-8093 Zurich, Switzerland}
\affiliation{Quantum Center, ETH Zurich, CH-8093 Zurich, Switzerland}

\author{Aleksandr Grigorev}
\affiliation{Department of Physics, ETH Zurich, CH-8093 Zurich, Switzerland}
\affiliation{Quantum Center, ETH Zurich, CH-8093 Zurich, Switzerland}

\author{Guillermo F. Pe\~{n}as}
\affiliation{Institute of Fundamental Physics (IFF), CSIC, Calle Serrano 113b, 28006 Madrid, Spain}

\author{Ricardo Puebla}
\affiliation{Department of Physics, Universidad Carlos III de Madrid, Avenida de la Universidad 30, 28911 Legan\'es, Spain}

\author{Juan Jos\'e Garc\'ia-Ripoll}
\affiliation{Institute of Fundamental Physics (IFF), CSIC, Calle Serrano 113b, 28006 Madrid, Spain}

\author{Jean-Claude Besse}
\affiliation{Department of Physics, ETH Zurich, CH-8093 Zurich, Switzerland}
\affiliation{Quantum Center, ETH Zurich, CH-8093 Zurich, Switzerland}

\author{Andreas Wallraff}
\affiliation{Department of Physics, ETH Zurich, CH-8093 Zurich, Switzerland}
\affiliation{Quantum Center, ETH Zurich, CH-8093 Zurich, Switzerland}

\author{Anatoly Kulikov}
\affiliation{Department of Physics, ETH Zurich, CH-8093 Zurich, Switzerland}
\affiliation{Quantum Center, ETH Zurich, CH-8093 Zurich, Switzerland}

\date{\today}

\begin{abstract}
    The tunable interaction between stationary quantum bits and propagating modes of light allows for the encoding of quantum information in the state of itinerant photons. This ability fulfills a central requirement for quantum networking, enabling quantum state transfer between distant quantum devices.
    Conventionally, a symmetric envelope of the photon wavepacket is used for such purposes. Yet, the use of alternative \textit{temporal modes}  enables multiple applications in waveguide quantum electrodynamics that remain unexplored experimentally.
    Here, we use superconducting quantum circuits to generate individual itinerant microwave photons shaped in three mutually orthogonal temporal modes. We transfer the created photons across a 30-m cryogenic link, showing that the orthogonality allows us to decide at the receiver which mode to absorb, reflecting the other two with a selectivity ratio of 40. This experimental capability extends the microwave-frequency quantum communication toolbox, enabling a new photonic degree of freedom.
        
\end{abstract}

\maketitle

The networking of quantum devices has attracted increasing interest in the past decades \cite{Kimble2008, Azuma2023, Knorzer2025}. Quantum communication has applications in the realm of secure information transfer and processing, such as quantum key distribution \cite{Nadlinger2022} and blind measurement-based quantum computation \cite{Walther2005, Wei2025}, and also plays a key role in the development of large quantum computing infrastructures. Scaling up the latter will require modularity, based on the connection of separate quantum devices in local and long-distance networks \cite{Niu2023, AghaeeRad2025, Main2025}.

Long-range quantum communication suffers from photon loss that scales exponentially with distance.
In this setting, probabilistic schemes are often used to improve the entanglement fidelity using post-selection \cite{Briegel1998, Barrett2005, LagoRivera2021, Pompili2021, Azuma2023, Krutyanskiy2023b}.
Alternatively, deterministic schemes are better suited for networks where losses are low and high-fidelity state transfer is feasible.
Deterministic state transfer is often executed between two nodes, one emitter and one receiver, which exchange single photons~\cite{Cirac1997} at optical~\cite{Ritter2012} or microwave frequencies~\cite{Kurpiers2018, Axline2018, Leung2019}.
This requires accurate control of the photon \textit{temporal mode} \cite{Raymer2020, Fabre2020}, as the receiver must adapt its absorption process to the envelope of the incident wavepacket.

Demonstrations of this deterministic protocol in short-range networks with superconducting circuits have so far been restricted to the use of simple bell-shaped envelopes \cite{Kurpiers2018, Qiu2025a} or slight numerical corrections of these that improve the photon transfer efficiency \cite{Almanakly2025}. But the high RF-tunability of Hamiltonian parameters offered by circuit QED systems \cite{Blais2021} enables the shaping of photons into alternative temporal modes \cite{Pechal2014, Penas2024}, whose applications in quantum communication are worth exploring experimentally \cite{Brecht2015b, Ansari2017a, Morin2019, Penas2023, Miyamura2025}.

In this work, we demonstrate primitives of quantum communication with a set of orthogonal temporal modes of microwave photons \cite{Pechal2014, Penas2024}. We generate the photons using a stationary transmon emitter, monitoring the population of its states during the emission process. We find that the desired photon envelopes are accurately produced. Furthermore, we show that, at the receiving node, we can choose to either absorb or reflect the incident excitation with a high selectivity, making use of the orthogonality of their mode functions \cite{Penas2024}, and without the transfer efficiency being significantly degraded for higher modes.

\begin{figure}[hbtp]
    \centering
    \includegraphics[width=\linewidth]{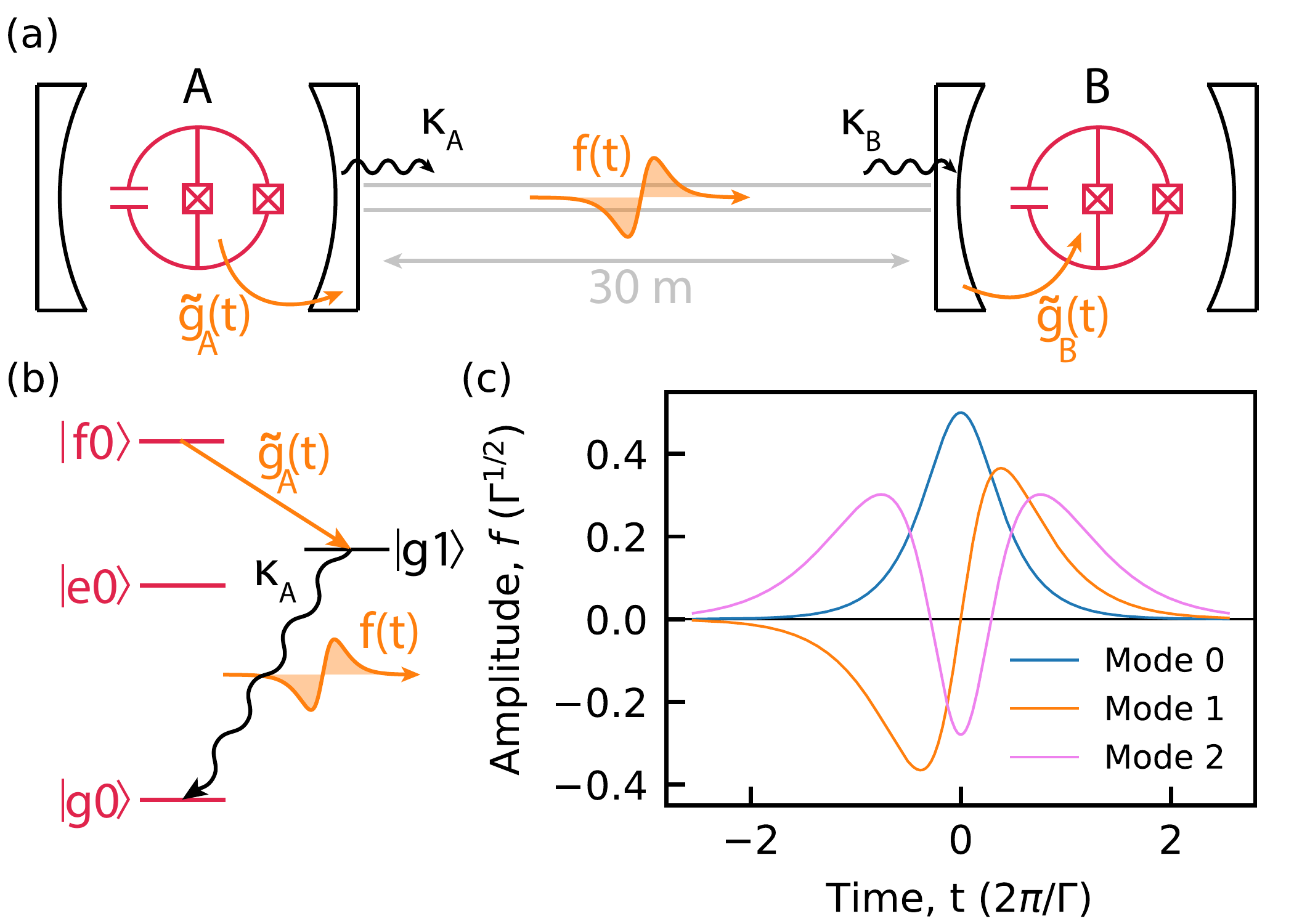}
    \caption{Experimental setup and orthogonal temporal modes.
    (a)~Communication setup. The transmon qubits (red) are coupled to dedicated resonators (black) at a rate $\gtildeAB(t)$, which are used to emit or absorb photons (orange) into/from a communication channel (gray) at a maximum rate $\kappaAB$.
    (b)~Energy levels of the joint transmon-resonator (red-black) system that are relevant for photon emission. A tunable coupling $\gtildeA(t)$ (orange arrow) between the $\ket{\rm f0}$ and $\ket{\rm g1}$ states and the subsequent decay of the resonator into the communication channel at rate $\kappaA$ (black arrow) produce a shaped photon (orange envelope).    
    (c)~Wavefunction amplitude versus time for the first three temporal modes in the orthogonal set, see Eqs.~(\ref{eq:f0}-\ref{eq:f2}). Both axes are normalized to the photon bandwidth parameter $\Gamma/2\pi = 14$~MHz selected for all three envelopes.}
    \label{fig:setup_modes}
\end{figure}

The experimental setup we use to emit and transfer photons consists of two nominally identical superconducting quantum devices, A and B, connected by a 30-m-long bidirectional link \cite{Magnard2020, Storz2023}, constituting a two-node network. The communication channel is a cryogenic waveguide that allows for low-loss exchange of microwave photons \cite{Kurpiers2017}, see \figref{fig:setup_modes}[a]. Each of the devices contains one flux-tunable transmon qubit \cite{Koch2007}, whose $\ket{\rm g}$, $\ket{\rm e}$ and $\ket{\rm f}$ states can be controlled coherently with a charge line and read out via a coplanar resonator and a Purcell filter \cite{Walter2017, Swiadek2024}. A second resonator-filter pair mediates the interaction between the qubit and the communication channel, see \cite{Storz2023} for more details on the devices and the setup.

The central requirement for quantum communication is the controlled emission of a qubit excitation into the channel. At node A we emit a photon using a microwave-driven coupling between the joint transmon-resonator states $\ket{\rm f0}$ and $\ket{\rm g1}$, at rate $\gtildeA$. Simultaneously, the engineered coupling of the resonator to the channel, at an effective rate $\kappaA$, releases an itinerant photon at the resonator frequency into the waveguide, see \figref{fig:setup_modes}[b]. The envelope of the generated photon can be shaped into a temporal mode $f(t)$ by using an amplitude-and-phase-modulated drive pulse that realizes a tailored time-dependent transfer rate $\gtildeA(t)$ into the emission resonator \cite{Pechal2014, Kurpiers2018}.

In principle, in this setting we can generate any mode that does not require emission at a rate larger than $\kappaA$, that is $|f(t)|^2 \leq \kappaA\left[1-F(t)\right]$, where $F(t) = \int_{-\infty}^t|f(\tau)|^2d\tau$ is the cumulative photon emission probability \cite{Pechal2014, Zeytinoglu2015, Kurpiers2018, Penas2024}.
Here, time-symmetric modes are particularly suited for photon transfer, as a receiver device identical to the source device can absorb the emitted photons by time-reverting the coupling $\gtildeA(t)$ used for their generation \cite{Cirac1997}.
A symmetric photon temporal mode function that is frequently used in these settings \cite{Kurpiers2018, Miyamura2025, Almanakly2025} is
\begin{equation}\label{eq:f0}
    f_0(t) = \frac{\sqrt{\Gamma}}{2}
    \sech\left(\frac{\Gamma t}{2}\right),
\end{equation}
see the blue curve in \figref{fig:setup_modes}[c]. Beyond this fundamental mode, here we experimentally realize the next two higher-order envelopes in the orthogonal set proposed in Ref.~\cite{Penas2024},
\begin{align}
    f_1(t) &= \frac{\sqrt{3\Gamma}}{2\pi}\Gamma t\cdot\sech\left(\frac{\Gamma t}{2}\right)\label{eq:f1}\\
    f_2(t) &= \frac{\sqrt{5\Gamma}}{8\pi^2}\left(3\Gamma^2t^2-\pi^2\right)\cdot\sech\left(\frac{\Gamma t}{2}\right),\label{eq:f2}
\end{align}
which take the shapes shown for a fixed value of $\Gamma$ in orange and magenta in \figref{fig:setup_modes}[c].
These and higher-order modes are given by polynomial orthogonality under the weight function $\sech^2\left(\Gamma t/2\right)$, and are computed using the Gram-Schmidt method~\cite{Penas2024}, see App.~\ref{app:orthogonality}. The index $n$ corresponds to the number of nodes in the propagating temporal wavefunction $f_n(t)$.

\begin{figure}[hbtp]
    \centering
    \includegraphics[width=\linewidth]{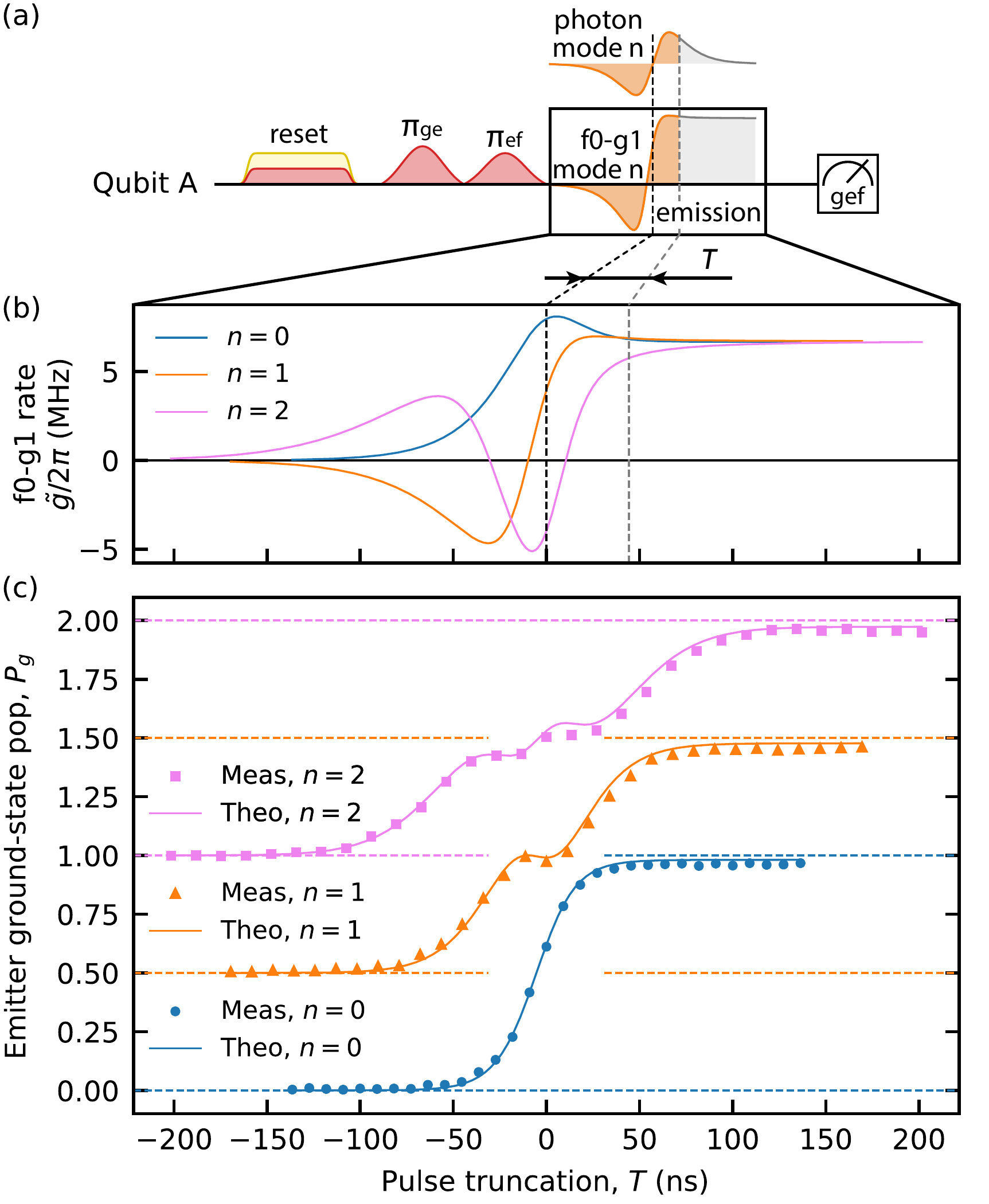}
    \caption{Photon emission in three orthogonal temporal modes.
    (a)~Gate sequence used to generate an itinerant photon while tracking the population of the source qubit. Microwave drives addressing the qubit transitions (red) and the f0-g1 reset transition to the readout resonator (yellow) are used to prepare state $\ket{\rm f}$. Then, the shaped f0-g1 pulse (orange) produces a photon with the desired envelope (orange, above), both being truncated at time $T$, after which the emitter transmon is read out.
    (b)~Calculated time-dependent Rabi rates applied to the f0-g1 transition to emit modes (\ref{eq:f0}, \ref{eq:f1}, \ref{eq:f2}), using $\Gamma/2\pi = 14$~MHz and $\kappaA/2\pi = 26.7$~MHz.
    (c)~Ground-state population of the qubit versus truncation time $T$ (referenced to the center of the target mode) for emission into modes $f_0$ (blue), $f_1$ (orange) and $f_2$ (magenta). The markers are measured data and the solid lines show the theory. The three datasets are offset by 0.5 for clarity, as indicated by the dashed lines.}
    \label{fig:emission}
\end{figure}

The protocol for the generation of a photon in one of the modes
(\ref{eq:f0}, \ref{eq:f1}, \ref{eq:f2})
starts with the reset of the emitter qubit to the $\ket{\rm g}$ state using an f0-g1 transition into the readout resonator 
\cite{Magnard2018}, and its subsequent excitation to the $\ket{\rm f}$ state \cite{Motzoi2009}. Then, we execute the emission operation using the f0-g1 transition into the transfer resonator as explained above. Finally, we perform dispersive readout \cite{Walter2017, Swiadek2024} of the source qubit to find the population of the states $\{\ket{\rm g}, \ket{\rm e}, \ket{\rm f}\}$, see pulse sequence in \figref{fig:emission}[a]. The time-dependent f0-g1 rate
\begin{equation}\label{eq:gtilde}
    \gtildeA(t) = \frac{\dot{f}(t) + \frac{\kappaA}{2}f(t)}{\sqrt{\kappaA[1-F(t)]-f(t)^2}},
\end{equation}
that is required to emit each mode is real-valued and can be computed from its target shape $f(t)$ and the resonator decay rate $\kappaA$, see calculated rates in \figref{fig:emission}[b] and App.~\ref{app:gtilde}.
Note that the shaped pulse that enacts this coupling $\gtildeA(t)$ must have a finite duration; in practice we trim its ends with 3-ns-long linear tapers,
ensuring that at least 99\% of the photon envelope remains within the unmodified time window.

We characterize the shapes of the generated photons indirectly, by tracking the qubit population during the emission process. The direct measurement of the quadrature amplitudes of the produced photons
\cite{Eichler2012} is not possible in the used setup since the bidirectional channel is not interrupted by a circulator to extract the photons. For this purpose, we truncate the f0-g1 emission pulse prematurely at a variable time $T$ as shown in \figref{fig:emission}[a,b]. The subsequent readout results capture the population dynamics of the qubit throughout the photon emission operation.

We observe that the ground-state population $P_{\rm g}(T)$ evolves from zero to one as the system is coherently driven from $\ket{\rm f}$ to $\ket{\rm g}$ by the f0-g1 pulse, thereby releasing a photon, see \figref{fig:emission}[c].
The characteristic duration of this operation is inversely proportional to the photon bandwidth $\Gamma/2\pi = 14$~MHz and increases slightly with the mode order $n$.
One distinctive feature of the population dynamics observed for the different modes is the presence of $n$ plateaus during the emission of mode $n$.
These flat intervals correspond to the times at which the drive amplitude, and therefore the instantaneous emission rate, evaluate to zero and vanish momentarily, creating the nodes in the photon wavefunction.

The calculated theory curves follow the measured population dynamics for the target modes $f_n(t)$ with no free fit parameters, see \figref{fig:emission}[c] and App.~\ref{app:qubit_population_em}. The resonator linewidth $\kappaA$ and the decay time $T_1^{\rm ef}$ in the e-f subspace of the transmon have both been measured independently. The good agreement between the measured data and the model, together with the observed population plateaus thus constitute an unequivocal signature that photons are shaped as expected \cite{Penas2024}.

Photons shaped into different modes are a resource with multiple applications: for example, it has been proposed to encode quantum information in orthogonal temporal modes and realize mode-selective gates, constituting a complete framework for all-photonic universal quantum computation \cite{Brecht2015b}.
Of particular interest is the use of photons in orthogonal modes for quantum communication, which we demonstrate next.

In communication experiments using the fundamental mode $f_0$, which is symmetric, the receiver typically executes the time-inverse of the operation used for emission, which absorbs the incoming photon into the $\ket{\rm f0}$ state of its memory qubit deterministically \cite{Kurpiers2018}.
Here we extend this method by transferring a single photon between the qubits located at the two nodes of our network using the three orthogonal photon envelopes discussed above, and experimentally probe their orthogonality. 

\begin{figure}[hbtp]
    \centering
    \includegraphics[width=\linewidth]{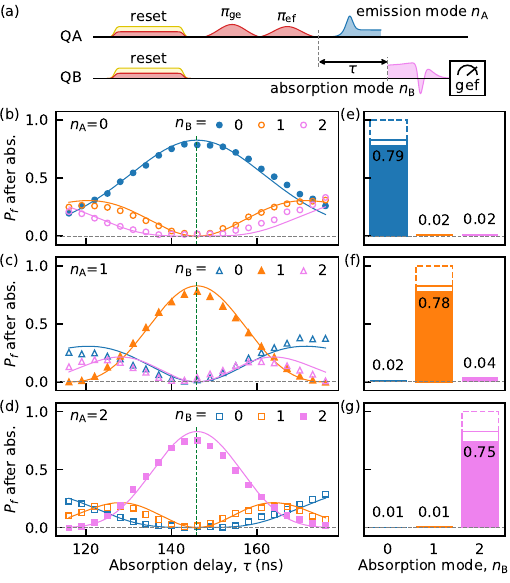}
    \caption{
    Photon transfer and mode-selective absorption.
    (a)~Pulse sequence used to characterize the mode orthogonality. After initialization of the qubits A and B in states $\ket{\rm f}$ and $\ket{\rm g}$ respectively (red and yellow), we apply the f0-g1 emission pulse of mode $\nA$ (blue) to qubit A and the absorption pulse of mode $\nB$ (pink) to qubit B, with a delay $\tau$. Finally, we read out the receiver transmon, B.
    (b, c, d)~Measured $\ket{\rm f}$-state populations in qubit B, when using the indicated modes for the emission and absorption pulses ($\nA, \nB$). Measured data is represented with markers, whose shape indicates the emission mode $\nA$ and color  indicates the absorption mode $\nB$ (filled when $\nA=\nB$). The solid lines are fits of the data to the model.
    The green dashed line indicates the optimal absorption delay $\tau_0$ extracted from the fit. The grey dashed line indicates zero population.
    (e, f, g)~Population of the $\ket{\rm f}$-state after absorption with the optimal delay $\tau_0$, using mode $\nA=0,1,2$ respectively for emission, and the indicated mode $\nB$ for absorption. Empty bars with solid and dashed frames represent the fitted model and the ideal values, respectively.
    }
    \label{fig:communication}
\end{figure}

The transfer experiment starts with the generation of a photon in mode $\nA$ from qubit A as explained above. We then apply the absorption pulse at qubit B, tailored for mode $\nB$ and delayed by $\tau$ from the emission operation to account for the propagation time along the waveguide. To compensate for small parameter differences between the nominally identical devices at the two nodes, which break the symmetry between emission and absorption, we adjust the two f0-g1 pulses as explained in App.~\ref{app:absorption}. Finally, we measure the receiver qubit $\ket{\rm f}$-state population, $P_{\rm f}$, to characterize the fidelity of the process, see gate sequence in \figref{fig:communication}[a].

We sweep the delay $\tau$ and repeat the experiment for emission and absorption mode indices $\nA$ and $\nB$ ranging from 0 to 2 each, always using $\Gamma/2\pi = 24$~MHz, and evaluate the transfer efficiency, which is approximated by the receiver qubit population $P_{\rm f}$, for each combination. We observe that the itinerant photon is transferred to the memory qubit in the receiver node when we use the same mode for both operations $(\nA=\nB)$. This is indicated by the high transfer efficiency of above 75\%. This efficiency is maximized around an optimal delay $\tau_0 = 145.9\pm0.2~{\rm ns}$ consistent with the group velocity $v_{\rm g}\approx0.72c$, where $c$ is the speed of light in vacuum, and length $L\approx30~{\rm m}$ of the rectangular waveguide, considering a small additional delay that stems from approximately 80~cm of coaxial cables that connect the waveguide to the samples, see the filled markers in \figref{fig:communication}[b--d]. Away from the optimal delay $\tau_0$, the measured population decreases gradually. This decrease is slower for temporal mode function $f_0$, as compared to modes $f_1$ and $f_2$, but always comparable to the inverse of $\Gamma$.

Conversely, when using different modes for emission and absorption ($\nA\neq \nB$, empty markers), the final $\ket{\rm f}$-state population of qubit B after the receiver pulse is approximately zero around the optimal delay.
This indicates that the incident photon is reflected back into the channel,
and shows that the generated temporal modes are in good approximation orthogonal.
Furthermore, the minima in the $0\leftrightarrow2$ cases are wider than those for $0\leftrightarrow1$ and $1\leftrightarrow2$, indicating that the orthogonality between modes $f_0$ and $f_2$ is more robust to an imperfect calibration of the absorption delay.

To quantify the dominant error mechanisms in the photon transfer process, we compare the measured data to a model, where the receiver qubit population $P_{\rm f}$ is given by the (delayed) overlap between the temporal mode used for the absorption pulse and the incident mode, see solid lines in \figref{fig:communication}[b--d] and derivation in App.~\ref{app:orthogonality}. Fitting this model to the measured data using only two global parameters, we extract the optimal absorption delay $\tau_0$ (see above) and the photon loss probability $p_{\rm loss} = 17\%$. We find that all the combinations of emission and absorption modes $(\nA, \nB)$ agree well with our model.
The photon loss is consistent with the one observed in earlier experiments~\cite{Storz2023, Kulikov2024} performed only with mode $f_0$. This loss predominantly originates from resistivity and impedance mismatches at the interfaces between the chips and the communication channel, errors in the control pulses and decoherence of the qubits. The loss of the waveguide itself is negligible, with a measured attenuation below 1~dB/km \cite{Magnard2020}. In our model, we do not consider wavepacket distortions in the channel and photon shaping errors, which could explain the deviations between the calculations and the experimental data.

Finally we demonstrate a direct application of orthogonal temporal modes to photon transfer, by fixing the absorption delay to the optimal value $\tau_0$ and analyzing the optimal photon transfer efficiencies. When the two modes are equal $(\nA=\nB)$, the transfer efficiencies are between 79\% and 75\%, decreasing weakly with increasing mode order. When the modes are different $(\nA\neq \nB)$, the reflection of the photon leaves the receiver qubit in the ground state, with populations $P_{\rm f}$ between 1\%
and 4\%, see filled bars in \figref{fig:communication}[e--g].
These nine measured populations constitute a transfer matrix $M_{\nA,\nB} = P_{\rm f}^{\nA,\nB}(\tau_0)$ which is approximately diagonal. We quantify this with the \textit{selectivity ratio} \cite{Besser2003} between its average diagonal and off-diagonal elements: $\Sigma = \overline{M_{\nA=\nB}}/\overline{ M_{\nA\neq \nB}} \approx 40 \approx 16~{\rm dB}$.
These results demonstrate that we have the freedom to choose which temporal mode to absorb at the receiver, thereby selectively rejecting the incoming photon if its mode is orthogonal.

In this work, we have experimentally investigated the generation, transmission and absorption of orthogonal temporal modes of microwave photons in a prototypical two-node network of superconducting qubits spanning a distance of 30 metres. 
The extension of the fundamental mode $f_0$ to the higher-order functions $f_1$ and $f_2$ comes at no major cost in the photon emission rate.
Exploiting the temporal mode degree of freedom upon absorption, we have demonstrated that the receiver can not only deterministically absorb the impinging microwave photon, but also selectively reflect it by simply modulating a control signal, with no additional requirements to the device architecture.

These results showcase the high degree of controllability of purely-microwave quantum communication networks and demonstrate a useful primitive for quantum networking, enabling orthogonal-mode encoding of photonic qubits for the detection and correction of photon loss \cite{Borregaard2020, Bell2023}, as an alternative to the previously-existing frequency encoding \cite{Yang2025a, Miyamura2025, Wang2025s} and time-bin encoding \cite{Kurpiers2019}. Although the direct application of the methods explored here to multiplexed communication seems challenging at the moment \cite{Penas2024}, they could be used to route photonic qubits across multiple nodes connected to a shared bus \cite{Almanakly2025}, where the mode would encode the target address \cite{Dally2004}.
However, with the presented temporal modes, theory indicates that the wavepacket gets distorted upon reflection from the receiver $(n\neq m)$ by the application of a mismatched absorption pulse \cite{Penas2024}. To address this obstacle, further theoretical investigation on the distortion of the realized mode functions or alternative ones remains necessary.

As a direct application, the family of modes we have used in the present work can be extended to span a basis for the tomography of radiation in the communication channel. Our results enable the use of qubit measurements for {\it in-situ} characterization of photon shapes and their distortions in closed, bidirectional networks \cite{Brecht2015b, Ansari2017a, Morin2020, Penas2022, Reuer2022}. When aiming to correct for photon distortions, orthogonal temporal modes may also constitute an efficient decomposition for optimal control \cite{Penas2023, Almanakly2025}, which may be advantageous for the development of high-fidelity deterministic quantum communication alongside low-loss channels and efficient qubit-photon interfaces.

Related experimental work, carried out concurrently and independently, is discussed in Ref. \cite{Sunada2026}

\section*{Acknowledgements}
The authors would like to thank J.~O'Sullivan, J.~Knörzer and X.~Dai for fruitful discussions and feedback on the manuscript and the figures. This work was supported by ETH Zürich. Additional support is acknowledged from the Spanish Government via the project PID2024-161371NB-C21 (MCIU/AEI/FEDER, EU), and the Ram\'on y Cajal (RYC2023-044095-I) research fellowship.

\section*{Author contributions}
A.H.A., A.K., J.D.S. and A.G. developed the measurement routines, using theory provided by G.F.P., R.P. and J.J.G.R.. A.K., J.S., A.G. and A.H.A. calibrated the experimental setup. A.H.A. and A.K. carried out the experiment and analyzed the data. A.H.A. and A.K. wrote the manuscript with input from all co-authors. J.J.G.R, J.C.B., A.W. and A.K. supervised the project.

\appendix

\section{Orthogonality of temporal modes}\label{app:orthogonality}

\subsection{Definition of orthogonality and mode functions}

The temporal modes used in this work are derived from the hyperbolic-secant function
\begin{equation}
    f_0(t) = \frac{\sqrt{\Gamma}}{2}
    \sech\left(\frac{\Gamma t}{2}\right).
\end{equation}
This wavefunction represents the probability amplitude of detecting the itinerant photon at time $t$. Starting from this mode one can define a set of real-valued functions $\{f_n(t)\}$ by multiplying the hyperbolic secant by polynomials of order $n$, such that the resulting envelopes form an orthogonal set under an overlap integral
\begin{equation}\label{eq:inner}
    ( f_n | f_m ) = \int_{-\infty}^{\infty} f_n^*(t)\ f_m(t) ~dt = \delta_{nm}.
\end{equation}

Based on the first mode and this inner product, we use the Gram-Schmidt method and obtain the next two mode functions in the orthogonal set:
\begin{align}
    f_1(t) &= \frac{\sqrt{3\Gamma}}{2\pi}\Gamma t\cdot\sech\left(\frac{\Gamma t}{2}\right),\\
    f_2(t) &= \frac{\sqrt{5\Gamma}}{8\pi^2}\left(3\Gamma^2t^2-\pi^2\right)\cdot\sech\left(\frac{\Gamma t}{2}\right).
\end{align}
These are the functions describing the modes that we have produced experimentally.

\subsection{Absorption of orthogonal modes}

In our experiment, microwave photons shaped into these modes are transferred between two devices. The emitter node sends a photon shaped in mode $f_\nA(t)$. Assuming the distortions in the channel are negligible, the mode that arrives at the receiver is $\sqrt{1-p_{\rm loss}} f_\nA(t)$, damped by photon loss. At the receiver, we apply a f0-g1 control $\gtilde_\nB(t)$, realizing a unitary evolution $U_\nB$ which is the inverse of the process that would produce a photon with envelope $f_\nB(t)$, that is:
\begin{equation}
	U_\nB^{-1}\left[\ket{\rm f}\otimes\ket{0}\right]
	 = \ket{\rm g}\otimes f_\nB
\end{equation}
Conversely, the absorption process
\begin{equation}
	U_\nB\left[\ket{\rm g}\otimes f_\nB\right] = \ket{\rm f}\otimes\ket{0}
\end{equation}
maps the temporal mode $f_\nB$ to the receiver $\ket{\rm f}$ state
However, the receiver may use an absorption pulse that does not match the incident temporal mode, that is $\nB\neq\nA$. Considering this possibility, we decompose the emitted mode $f_\nA$ as the linear combination
\begin{equation}
	f_\nA = \alpha f_\nB + \beta f^{\perp},
\end{equation}
where
\begin{align}
	\alpha &= ( f_\nB | f_\nA ),\\
	\beta &= ( f^{\perp} | f_\nA ),
\end{align}
and $f^{\perp}$ represents any contribution to the envelope $f_\nA$ that is orthogonal to $f_\nB$: $( f^{\perp}|f_\nB ) = 0$.
The absorption of the incoming excitation, including photon loss, thus produces the final state
\begin{align}
	\ket{\psi_{\rm final}} =& U_\nB\left[\ket{\rm g}\otimes \sqrt{1-p_{\rm loss}}f_\nA\right] = \nonumber\\
	=& \sqrt{1-p_{\rm loss}}\alpha U_\nB\left[\ket{\rm g}\otimes f_\nB\right] + \nonumber \\
	+&\sqrt{1-p_{\rm loss}}\beta U_\nB\left[\ket{\rm g}\otimes f^{\perp}\right] = \nonumber \\
	=& \sqrt{1-p_{\rm loss}}\left[\alpha \ket{\rm f}\otimes \ket{0} + \beta U_\nB\left(\ket{\rm g}\otimes f^{\perp}\right)\right].
\end{align}
The second term in this expression must remain orthogonal to $\ket{\rm f}\otimes \ket{0}$ and preserve the excitation number, which reduces the only possible result to
\begin{equation}
	\ket{\psi_{\rm final}} = \sqrt{1-p_{\rm loss}}\left[\alpha \ket{\rm f}\otimes \ket{0} + \beta \ket{\rm g}\otimes f_{\rm out}\right],
\end{equation}
where $f_{\rm out}$ is the temporal mode of the reflected photon in the case of a mismatched absorption pulse ($\nB\neq\nA$).
After the absorption operation, the $\ket{\rm f}$-state probability amplitude $c_{\rm f}$ of the receiver qubit is
\begin{align}\label{eq:abs_amplitude}
	c_{\rm f} &= \sqrt{1-p_{\rm loss}}\alpha = \sqrt{1-p_{\rm loss}}( f_\nB | f_\nA )  = \nonumber \\
	&= \sqrt{1-p_{\rm loss}}\int_{-\infty}^{\infty} f_\nB^*(t) f_\nA(t) dt.
\end{align}
This result allows us to better understand the effect of the used control field in the absorption process, resulting in two possible scenarios when the photon is not lost in the channel. First, according to this model, when the f0-g1 pulse matches the incident mode, the photon is perfectly absorbed, as previously implied by time-reversal symmetry. But second and foremost, this shows that the rejection mechanism observed experimentally follows directly from mode orthogonality. If the absorption drive aims at a mode orthogonal to the incoming one, the receiver qubit ends in the ground-state and the photon is not absorbed, but reflected back into the communication channel. For the modes used in this work, simulation indicates that the shape of this output photon $f_{\rm out}$ differs significantly from the input one \cite{Penas2024}.

\subsection{Dependence of the receiver qubit population on the absorption delay}
In our setup, we execute the absorption pulse with a delay $\tau$ with respect to the start of the emission pulse, aiming to find the optimal absorption time that accounts for the propagation of the photon across the link. The final population of the receiver qubit depends on this delay, as shown in Fig.~\ref{fig:communication}. Mathematically, the delay shifts the photon envelopes in (\ref{eq:abs_amplitude}) with respect to each other. Thus, the final state of the transmon can be derived from (\ref{eq:abs_amplitude}) and depends on the delay of the absorption pulse as the convolution
\begin{equation}\label{eq:cftau}
    c_{\rm f}(\tau) = \sqrt{1-p_{\rm loss}}\int_{-\infty}^{\infty}f_\nB^*(t-\tau)f_\nA(t)dt
\end{equation}
between the incident mode and the absorption mode. Consequently, the receiver qubit population after absorption is
\begin{align}\label{eq:Pftau}
	P_{\rm f}(\tau) &= |c_{\rm f}(\tau)|^2 = \nonumber \\
	&= (1-p_{\rm loss})\left|\int_{-\infty}^{\infty}f_\nB^*(t-\tau)f_\nA(t)dt\right|^2.
\end{align}
This is the magnitude measured in the photon transfer experiment in Fig.~\ref{fig:communication}, in which we realize a projective measurement of the incoming wavepacket. We use Eq.~\eqref{eq:Pftau} to analyze the data, plotting the results of this calculation as solid lines in Figs.~\ref{fig:communication}b,~\ref{fig:communication}c~and~~\ref{fig:communication}d, in which $p_{\rm loss}$ and $\tau$ have been used as fit parameters.

\section{Controlling the time-dependent qubit-resonator coupling rate $\gtilde(t)$\label{app:gtilde}}
\subsection{General derivation of $\gtilde(t)$ for the emission of photons with a mode function $f(t)$}
For the emission of photons with a temporal mode function $f(t)$ we use a time-dependent coupling rate $\gtilde(t)$ between the $\ket{\rm f0}$ and $\ket{\rm g1}$ states of the transmon-resonator system, as given in Eq.~\eqref{eq:gtilde}. Here we derive an expression for this time-dependent coupling rate, closely following~\cite{Penas2023,Penas2024}, for the emission of a photon with the same frequency as the transfer resonator. In the rotating frame of a resonant f0-g1 interaction and in the absence of decoherence, the system evolves within the f0-g1 subspace as
\begin{equation}\label{eq:ket_f0g1}
    \ket{\psi_{{\rm f0-g1}}(t)} = \alpha(t)\ket{\rm f0}-i\beta(t)\ket{\rm g1},
\end{equation}
where the amplitudes $\alpha(t)$ and $\beta(t)$ obey
\begin{align}\label{eq:app_alphadot}
   \dot{\alpha}(t)&=\gtilde(t)\beta(t),\\ \label{eq:app_betadot} \dot{\beta}(t)&=-\gtilde^*(t)\alpha(t)-\kappaA \beta(t)/2,
\end{align}
with $\kappaA$ being the coupling rate of the emission resonator of the source qubit.

For real-valued temporal modes of the output photon $f(t)=\sqrt{\kappaA}\beta(t)$, we consider $\alpha(t),\beta(t),\gtilde(t)\in\mathbb{R}$. We combine Eqs.~\eqref{eq:app_alphadot}-\eqref{eq:app_betadot}, finding $\frac{d}{dt}(\alpha(t)^2+\beta(t)^2)=-\kappaA \beta(t)^2$, which upon integration leads to
\begin{align}
    \alpha(t)^2=\alpha(t_0)^2+\beta(t_0)^2-\beta(t)^2-\int_{t_0}^t d\tau \kappaA \beta(t)^2.
\end{align}
Considering $\alpha(t_0)=1$ and $\beta(t_0)=0$ at time time $t_0$, we find that the population of the $\ket{\rm f0}$ state depends on the mode amplitude $f(t)$ and its cumulative probability $F(t)=\int_{t_0}^t d\tau f(\tau)^2$ as
\begin{align}\label{eq:alpha2_f0g1}
    \alpha(t)^2=1-\frac{f(t)^2}{\kappaA}-F(t).
\end{align}
As discussed in the main text, the experimental implementation of the emission pulse must have a finite duration. Therefore, we trim its ends with linear tapers on a time scale shorter than the inverse of the photon bandwidth $\Gamma$, ensuring that, ideally, at least 99\% of the photon envelope remains within the unmodified time window for all three modes. Thus, in this calculation we assume that the lower limit of the integral in $F(t)$ can be replaced by $-\infty$, for analytical simplicity.

The control rate $\gtilde(t)$ follows from Eq.~\eqref{eq:app_alphadot}, as
\begin{align}
    \gtilde(t)=\frac{\dot{\alpha}(t)}{\beta(t)}=\pm \sqrt{\kappaA}\frac{1}{f(t)}\frac{\frac{d}{dt}\left[1-f(t)^2/\kappaA-F(t) \right]}{2\sqrt{1-f(t)^2/\kappaA-F(t)}},
\end{align}
which further simplifies to 
\begin{align}\label{eq:gtilde_mp}
    \gtilde(t)=\frac{\mp(\dot{f}(t)+\frac{\kappaA}{2} f(t))}{\sqrt{\kappaA[1-F(t)]-f(t)^2}}.
\end{align}
The sign stems from the square root in the real-valued amplitude $\alpha(t)$
and only introduces a global phase in the emitted photon. Taking the positive sign in Eq.~\eqref{eq:gtilde_mp} we arrive at Eq.~\eqref{eq:gtilde} of the main text. For the denominator to be real-valued, the desired temporal mode $f(t)$ must be such that $f(t)^2 \leq \kappaA [1 - F(t)]$, i.e. the emitted photon flux at time $t$ cannot be larger than the coupling rate of the resonator into the channel times the remaining excitation probability in the qubit-resonator system, at any time. Otherwise the desired mode function is not physically feasible.

This result is also applied to the absorption of an incoming photon with a given mode function $f(t)$, by realizing the time-reversed emission process as explained in the main text, replacing the coupling rate of the resonator $\kappaA$ by that of the receiver device, $\kappaB$, in Eq.~\eqref{eq:gtilde}.

\subsection{Specific expressions of $\gtilde(t)$ for the temporal modes realized in the experiment}
Here we provide the specific coupling rates $\gtilde(t)$ for each of the three orthogonal temporal mode functions realized here, $f_0(t)$, $f_1(t)$ and $f_2(t)$ (cf. Eqs.~\eqref{eq:f0}-\eqref{eq:f2}), whose functional form depends on the bandwidth parameter $\Gamma$.

First, we compute the time-derivatives of the mode functions
\begin{align}
    \dot{f}_0(t) &=-\frac{\Gamma^{3/2}}{4}{\rm sech}(\Gamma t/2)\tanh(\Gamma t/2), \\
    \dot{f}_1(t) &=\frac{\sqrt{3}\Gamma^{3/2}}{4\pi}{\rm sech}(\Gamma t/2)(2-\Gamma t\tanh(\Gamma t/2)), \\
    \dot{f}_2(t) &= \frac{\sqrt{5}\Gamma^{3/2}}{16\pi^2}{\rm sech}(\Gamma t/2)\nonumber \\&\quad\times (12\Gamma t+(\pi^2-3\Gamma^2 t^2)\tanh(\Gamma t/2)).
\end{align}
Then, we compute the analytic expressions of the cumulative photon emission probabilities $F_n(t) = \int_{-\infty}^{t}d\tau f_n(\tau)^2$ for the three modes
\begin{align}
    F_0(t)=(1+&e^{-\Gamma t})^{-1},\\
    F_1(t)=\frac{1}{2\pi^2}&\left( 2\pi^2-3x^2-12x\log(1+e^{-x})\nonumber \right. \\ &\left. +12 {\rm Li}_2(-e^{-x})+3x^2\tanh(x/2)\right),\\
    F_2(t)=\frac{1}{32\pi^4}&\left(27\pi^4+30\pi^2x^2-45x^4\right. \nonumber \\&+120x(\pi^2-3x^2)\log(1+e^{-x})\nonumber\\&+2160 [x {\rm Li}_{3}(-e^{-x})+{\rm Li}_4(-e^{-x})]\nonumber \\&+5(\pi^2-3x^2)^2\tanh(x/2)\nonumber\\
    &\left.-120(\pi^2-9x^2){\rm Li}_{2}(-e^{-x})\right),
\end{align}
with $x=\Gamma t$ and the polylogarithmic function ${\rm Li}_n(z)=\sum_{k=1}^\infty z^k k^{-n}$.

We finally combine these expressions following Eq.~\eqref{eq:gtilde}, and find
\begin{align}
    \label{eq:app_g0}
    \gtilde_0(t)=&\frac{\sqrt{\Gamma}(1+e^{\Gamma t})(\kappaA-\Gamma \tanh(\Gamma t/2))}{4\sqrt{\kappaA+e^{\Gamma t}(\kappaA-\Gamma)}} {\rm sech}(\Gamma t/2),\\
    \label{eq:app_g1}
    \gtilde_1(t)=&\frac{\Gamma^{3/2}}{2}{\rm sech}(x/2)(2+\kappaA t-x \tanh( x/2))\nonumber \\ &\times\left[2x\kappaA(x+4\log(1+e^{-x}))-8\kappaA{\rm Li}_{2}(-e^{-x})\right.\nonumber \\& -x^2{\rm sech}^2(x/2)(\Gamma+\kappaA\sinh(x))\left. \right]^{-1/2},\\
    \label{eq:app_g2}
    \gtilde_{2}(t)=&\frac{\sqrt{\Gamma}}{2}{\rm sech}(x/2)\left[12\Gamma x-\kappaA \pi^2+3\kappaA x^2\right.\nonumber \\ &\left.+\Gamma(\pi^2-3x^2)\tanh(x/2)\right]\nonumber \\
    &\times\left[48\kappaA x(3x^2-\pi^2)\log(1+e^{-x})\right. \nonumber\\&-864\kappaA (x{\rm Li}_3(-e^{-x})+{\rm Li}_4(-e^{-x}))\nonumber\\&+48\kappaA(\pi^2-9x^2){\rm Li}_{2}(-e^{-x})\nonumber\\&+(\pi^2-3x^2)^2(2\kappaA-\Gamma {\rm sech}^2(x/2))\nonumber\\&\left.-2\kappaA(\pi^2-3x^2)^2\tanh(x/2) \right]^{-1/2}.
\end{align}
These are the coupling rates implemented in the experiment, which are shown in \figref{fig:emission}[b].

\section{Qubit population during photon emission} \label{app:qubit_population_em}

The communication channel between the two nodes in our network is closed, which impedes the direct observation of the photons via heterodyne detection \cite{Eichler2012, Kurpiers2018}. Instead, the emission process has been characterized indirectly, by monitoring the population of the emitter qubit throughout the operation. Here we analyze the population dynamics observed in the experiment, see Fig.\ref{fig:emission}c.

During the emission of a single photon Fock state $\ket{1}$ with mode function $f(t)$ using an f0-g1 drive, and in the absence of decoherence, the qubit population is transferred from $\ket{\rm f}$ to $\ket{\rm g}$, following Eq.~(\ref{eq:ket_f0g1}). In this process, the f-state population of the emitter qubit at a time $T$ is given by Eq.~(\ref{eq:alpha2_f0g1}), as
\begin{equation}
    P_{\rm f}(T) = \alpha(T)^2 = 1 - \frac{f(T)^2}{\kappaA} - \int_{-\infty}^{T}f(t)^2 dt.
\end{equation}
Thus the ground-state population is
\begin{equation}\label{eq:qubit_pop_theo}
    P_{\rm g}(T) = 1 - P_{\rm f}(T) = \frac{f(T)^2}{\kappaA} + \int_{-\infty}^{T}f(t)^2 dt,
\end{equation}
which evolves from zero to one as the emitter excitation is driven into the communication channel. The probability of finding the qubit in the ground state is equal to that of the photon having been transferred coherently to the emission resonator or the waveguide. At time $T$ during the emission, the probability of finding the photon in the resonator is $\beta(T)^2 = f(T)^2/\kappaA$, and the probability that it has been released into the communication channel at any prior time is the integral of $f(t)^2$ for $t<T$. The sum of the two contributions results in Eq.~\eqref{eq:qubit_pop_theo}.

The main incoherent error in the emission process is decay in the source qubit from the $\ket{\rm f}$ to the $\ket{\rm e}$ state at a rate $1/T_1^{\rm ef}$. Consequently, the population of the emission resonator at time $T$ is reduced by a factor $e^{-T/T_1^{\rm ef}}$, whereas the photon emission probability at each prior time $t$, i.e. the integrand, is suppressed by a time-dependent factor $e^{-t/T_1^{\rm ef}}$. As a result, we introduce qubit damping in our model as
\begin{equation}\label{eq:qubit_pop_model}
    P_g(T) = \frac{|f(T)|^2}{\kappaA} e^{-T/T_1^{\rm ef}} + \int_{-\infty}^{T}|f(t)|^2 e^{-t/T_1^{\rm ef}} dt,
\end{equation}
This model, plotted as solid lines in Fig.~\ref{fig:emission}c, agrees well with the measured populations, without any fit parameters.

\section{Experimental details of photon absorption}\label{app:absorption}

In the presented experiment, the devices were designed to be identical, but small systematic variations in the fabrication and preparation of the samples give rise to small differences between the two. Wafer-scale variations in the metal thickness or etch depth of the base Nb layer yield slightly different resonance frequencies in the photon transfer resonators: $\omega_{\rm r}^{\rm B} \neq \omega_{\rm r}^{\rm A}$. Moreover, impedance mismatches in the emission line, due to wire-bonds or connectors on the PCB, lead to linewidths of these resonators differing slightly from each other: $\kappaA=26.7~{\rm MHz}$, $\kappaB=30.7~{\rm MHz}$.
Here we explain how we adjust the emission and absorption pulses to account for these asymmetries.

Compensating for the difference in resonator linewidths is straightforward.
To absorb the photon incident at the receiver, we apply the inverse of a drive pulse that would create a photon at the receiver with the temporal mode of the incoming photon.
To do so, we use temporal modes with $\Gamma<{\rm min}(\kappaA, \kappaB)$, which both nodes are capable of generating. Then, for the emission and absorption coupling rates $\gtildeAB(t)$, we use the same target photon envelope (same $\Gamma$, same $f(t)$), but the resonator linewidth $\kappaAB$ of the corresponding node, see Eq.~(\ref{eq:gtilde}).

\begin{figure}[!t]
    \centering
    \includegraphics[width=0.8\linewidth]{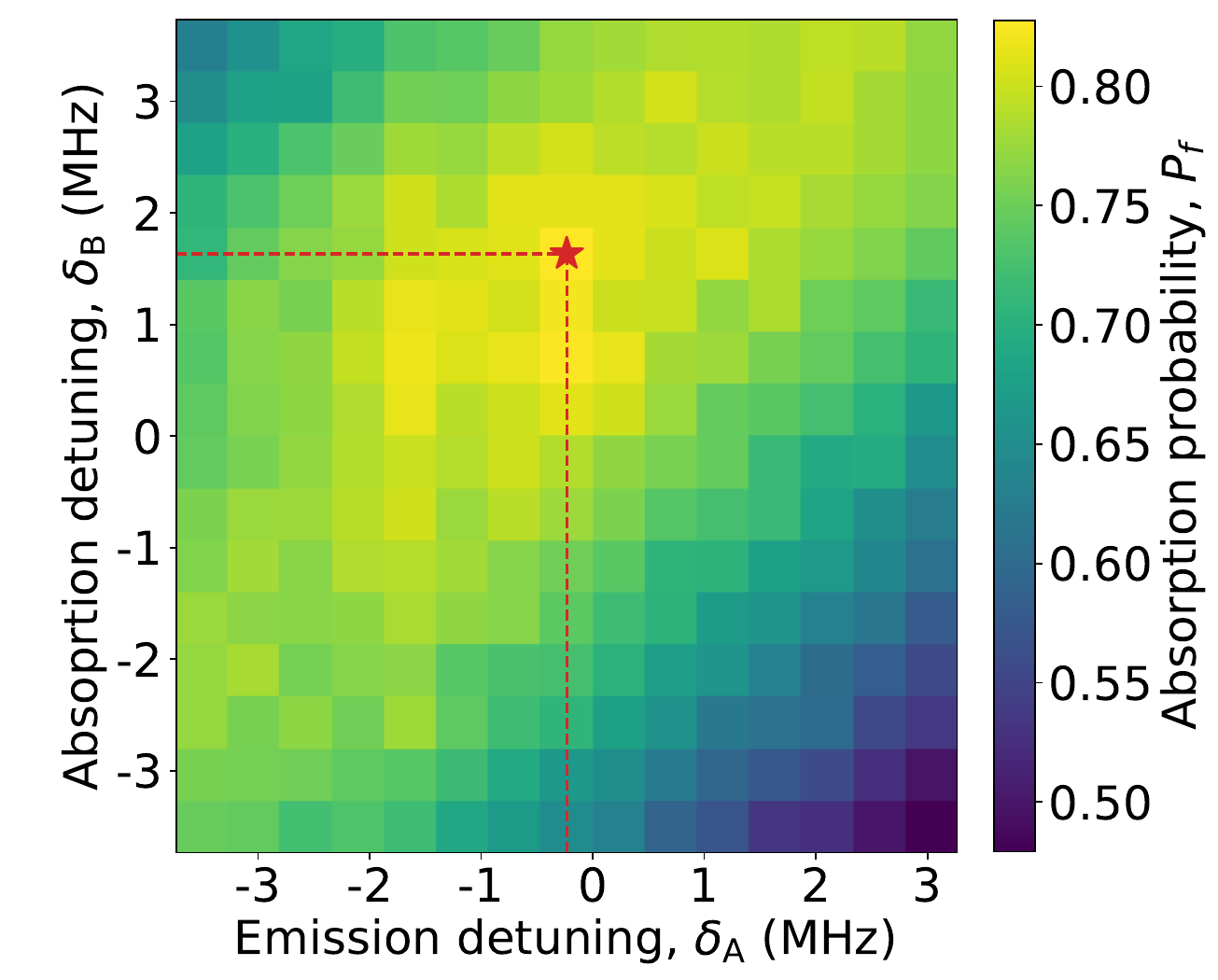}
    \caption{\textbf{Calibration of f0-g1 detuning.} Receiver qubit population $P_{\rm f}$ \emph{vs}. detuning $\deltaA$ and $\deltaB$ of the f0-g1 emission and absorption pulses. The maximal population (red marker) indicates the optimal detunings (red dashed lines).}
    \label{fig:app_stupid_detuning}
\end{figure}

The frequency misalignment of the transfer resonators $\DeltaAB/2\pi = (\omega_{\rm r}^{\rm B} - \omega_{\rm r}^{\rm A})/2\pi\approx 2~{\rm MHz}$ results in a frequency difference between the emitted and absorbed photons, which
impacts the transfer fidelity \cite{Qiu2025a} and needs to be addressed.
To compensate for the frequency difference between the emission and absorption resonator, we generate photons with a frequency $\omega_r+\delta$, detuned by $\delta$ from the resonator frequency $\omega_{\rm r}$, by shifting the carrier frequency of the pulse by $-\delta$, to $\omega_{\rm f0g1} = \omega_{\rm f} - \omega_{\rm r} - \delta$, as we are using the red f0-g1 sideband \cite{Magnard2018}. To characterize the effectiveness of this method, we sweep the detuning of both the emission ($\deltaA$) and absorption ($\deltaB$) pulses \cite{Campbell2026} and monitor the receiver qubit population after absorption, $P_{\rm f}$, in the photon transfer experiment sketched in Fig.~\ref{fig:communication}a, using mode $f_0$.

We find that the detunings resulting in the highest f-state population at the receiver qubit follow a trend that approximately meets $\deltaB-\deltaA =\DeltaAB$, see Fig~\ref{fig:app_stupid_detuning}. However, the population we measure is generally lower for larger absolute values of $\deltaAB$. For larger detunings the offset drive introduces unwanted distortions in the photon temporal mode, reducing its symmetry \cite{Miyamura2025}. Still, as the used detunings are much smaller than $\kappaAB$, these distortions are small. The interplay between energy conservation and photon mode distortions gives rise to a local maximum in the receiver population. From it, we pick $\deltaA/2\pi$ = -0.30 MHz and $\deltaB/2\pi$ = 1.90 MHz and use those values for the photon transfer experiments. With these parameters, both nodes emit photons at the same frequency. Detuning the drives by $\deltaAB$, we observe a receiver qubit population of $P_{\rm f}\approx83\%$, which is larger than the $79\%$ measured otherwise.

Alternative approaches to address frequency misalignment between emitter and receiver are discussed in the literature \cite{Miyamura2025, Pernas2026}.

%

\end{document}